# $^{40}$Ar/$^{39}$Ar ages of lunar impact glasses: Relationships among Ar diffusivity, chemical composition, shape, and size

N. E. B. Zellner[1] and J. W. Delano[2]

[1] Department of Physics, Albion College, Albion, MI 49224 USA
[2] New York Center for Astrobiology, Department of Atmospheric and Environmental Sciences, University at Albany (SUNY), Albany, NY 12222 USA

**Abstract**
Lunar impact glasses, which are quenched melts produced during cratering events on the Moon, have the potential to provide not only compositional information about both the local and regional geology of the Moon but also information about the impact flux over time. We present in this paper the results of 73 new $^{40}$Ar/$^{39}$Ar analyses of well-characterized, inclusion-free lunar impact glasses and demonstrate that size, shape, chemical composition, fraction of radiogenic $^{40}$Ar retained, and cosmic ray exposure (CRE) ages are important for $^{40}$Ar/$^{39}$Ar investigations of these samples. Specifically, analyses of lunar impact glasses from the Apollo 14, 16, and 17 landing sites indicate that retention of radiogenic $^{40}$Ar is a strong function of post-formation thermal history in the lunar regolith, size, and chemical composition. This is because the Ar diffusion coefficient (at a constant temperature) is estimated to decrease by ~3-4 orders of magnitude with an increasing fraction of non-bridging oxygens, X(NBO), over the compositional range of most lunar impact glasses with compositions from feldspathic to basaltic. Based on these relationships, lunar impact glasses with compositions and sizes sufficient to have retained ~90% of their radiogenic Ar during 750 Ma of cosmic ray exposure at time-integrated temperatures of up to 290K have been identified and are likely to have yielded reliable $^{40}$Ar/$^{39}$Ar ages of formation. Additionally, ~50% of the identified impact glass *spheres* have formation ages of ≤500 Ma, while ~75% of the identified lunar impact glass shards and spheres have ages of formation ≤2000 Ma. Higher thermal stresses in lunar impact glasses quenched from hyperliquidus temperatures are considered the likely cause of poor survival of impact glass spheres, as well as the decreasing frequency of lunar impact glasses in general with increasing age. The observed age-frequency distribution of lunar impact glasses may reflect two processes: (i) diminished preservation due to spontaneous shattering with age; and (ii) preservation of a remnant population of impact glasses from the tail end of the terminal lunar bombardment having $^{40}$Ar/$^{39}$Ar ages up to 3800 Ma. A protocol is described for selecting and analysing lunar impact glasses.

# 1. Introduction

The Moon provides the most complete history of impact events in the inner Solar System since its formation ~4500 million years ago (e.g., Fassett and Minton, 2013; Kirchoff et al., 2013; Morbidelli et al., 2012; LeFeuvre and Wieczorek, 2011; Stöffler et al., 2006; Neukum et al., 2001; Stöffler and Ryder, 2001). Since the Moon and Earth are close together in space, if properly interpreted, the Moon's impact record can be used to gain insights into how the Earth has been influenced by impacting events over billions of years. The timing of impacts on the Moon, however, is not well understood and is important for several reasons (NRC, 2007).

Since lunar impact glasses are droplets of melt produced by energetic cratering events and quenched during ballistic flight away from the target, their isotopic ages have the potential to provide constraints on the impact flux during the last several billion years, if the data are interpreted correctly. The impact flux can then be used to address the persistent question of whether or not there was a lunar cataclysm at around 3900 Ma (Tera et al., 1974) and what its relationship to the late heavy bombardment (LHB; e.g., Ryder et al., 2000) may be. Other questions about the impact flux can also be addressed. In addition, impact glasses sample widespread and random locations on the Moon making them a powerful tool for geochemical exploration of the Moon's crustal composition (Zellner et al., 2002; Delano, 1991), even though the location of impact ejection may not be known. Additionally, the compositions of glasses collected at a specific site can tell us about the geographic, and stratigraphic, character of that site, when well-established criteria for confidently distinguishing lunar impact-generated glasses from lunar volcanic glasses (Delano, 1986) are applied.

In the past decade or so, impact glasses have been increasingly used as tools to address the impact flux. Culler et al. (2000) studied 155 spherical glasses from the Apollo 14 landing site and interpreted the results in the context of both global lunar impacts and delivery of biomolecules to the Earth's surface. In particular, they interpreted their $^{40}Ar/^{39}Ar$ isotopic data on those glass spheres (without having attempted to distinguish between impact glasses and volcanic glasses) as evidence for (i) an increased impact flux around 3900 Ma (the purported "cataclysm") and (ii) a factor of 3.7 ± 1.2 increase in the last 400 Ma (Muller, 2002; Muller et al., 2001; Culler et al., 2000). In order to distinguish between impact and volcanic glasses, Levine et al. (2005) chemically analyzed the surfaces of spherical glasses from the Apollo 12 landing site and obtained $^{40}Ar/^{39}Ar$ ages on 81 lunar impact glasses. Although they also concluded that the age-distribution of their impact glass spheres was consistent with an apparent increase in the recent impact flux, Levine et al. (2005) suggested that local, young cratering events could be causing young spherical impact glasses to be disproportionately represented.

While interesting, these studies were incomplete in the following ways: (i) chemical compositions of the glasses were not determined (Culler et al., 2000), (ii) glasses of volcanic origin were not excluded from the data-set (Culler et al., 2000), and (iii) xenocryst-free, homogenous impact glasses were not solely used (Levine et al., 2005). Since Culler et al. (2000) did not provide descriptions of their glass spheres, item 'iii' may also apply to that investigation. The first and second concerns are important because it is not relevant to include the isotopic ages of lunar volcanic glasses when reporting an impact flux. For example, Delano (1988) reported that nearly 50% of the glasses in the youngest regolith breccia, 14307, studied at the Apollo 14 site (i.e., most similar to the current regolith) were of volcanic origin. In addition, since those volcanic glasses were more frequently spherical in shape than were the impact glasses, it is plausible that Culler et al. (2000) had a significant proportion of volcanic ages among their reported ages. The third concern is important because inherited Ar from undegassed crystalline inclusions can affect the reported $^{40}Ar/^{39}Ar$ formation age of a glass (Jourdan, 2012; Huneke et al., 1974), thereby contaminating the inferred age-distribution of lunar impact events. Finally, both groups assumed that each impact glass was formed in its own discrete impact event and thus that multiple glasses could not be formed in the same impact event.

We have obtained geochemical and chronological data on almost 100 xenocryst-free, homogeneous (or nearly so) impact glasses from the Apollo 14, 16, and 17 landing sites and with subsets of these ~100 samples, we have demonstrated the efficacy of interpreting these data together to understand the history of the sample(s). For example, Delano et al. (2007) showed that four glass shards (i.e., fragments, not spheres) with the same composition ('low-Mg high-K Fra Mauro' ('lmHKFM') glasses of Delano et al., 2007; 'basaltic-andesite' glasses of Korotev et al., 2010 and Zeigler et al., 2006) from the Apollo 16 landing site were formed at the same time, in one event (and not four). Therefore, the approach of interpreting the age data in the context of the compositional data allows for a better interpretation of the impact flux, so that it is not artificially inflated. This study additionally reported that spherical glasses are more likely to possess the local regolith composition, while non-spherical glasses (i.e., shards, fragments) are more likely to possess a non-local composition. Zellner et al. (2009a,b) combined geochemistry, age, and shape to interpret the ages and provenance of impact glasses from several Apollo landing sites. Impact ages of 12 individual glasses from the Apollo 17 landing site (Zellner et al., 2009a) revealed that only nine impact events may have been involved, depending on the compositional grouping selected. A clustering of $^{40}Ar/^{39}Ar$ ages at ~800 Ma (Zellner et al., 2009b) was observed in nine glasses from the Apollo 14, 16 and 17 landing sites, as well as in glasses from the Apollo 12 landing site (Levine et al., 2005), and at least seven separate impact events appear to have been involved in generating those glasses (Zellner et al., 2009b).

Glasses from the Apollo 16 landing site were investigated by Hui et al. (2010), who specifically selected low-K glasses, classified as spherules with various shapes, in order to address the local impact flux at the Apollo 16 landing site. About 130 glasses from a sample of Apollo 16 regolith were analysed for major and minor elements, and 30 of them (unpolished, to preserve sample-mass and the argon) had their $^{40}Ar/^{39}Ar$ ages determined. Some of those glasses appear to be neither homogeneous nor xenocryst-free (see Figure 3 in Hui et al., 2010). In order to distinguish among specific impact events, Hui et al. (2010) reported major- and minor-element compositions in addition to the $^{40}Ar/^{39}Ar$ ages for the impact glasses. Norman et al. (2012) suggested that in excess of 30% of glasses in a sample set could have been formed during the same impact event (i.e., glasses with the same composition and age). Even after accounting for multiple glasses formed in the same event, Hui et al. (2010) reported a high proportion of glasses (i.e., 'spherules') with ages <500 Ma, which they interpreted as being due to an increase in the recent impact flux (<500 Ma), though they reported that regolith dynamics or surface collection could also be a possible explanation. An important result of that detailed study was the observation that the exterior (i.e., the "rind") of the impact glass has a composition that is different from the bulk composition of the glass, which may become a useful constraint for inferring the provenance of a glass's origin, as described below.

Most recently, Norman et al. (2012) reported chemical compositions, $^{207}Pb/^{206}Pb$ model ages, and U-Th-Pb "chemical ages" for *spherical* glasses of volcanic and impact origins from the Apollo 17 landing site. The volcanic glasses had ages that were broadly consistent with those of known episodes of lunar mare volcanism. The impact glasses were compositionally similar to the local regolith, which consists largely of a mixture of highland rock and local mare basalts (as defined by Rhodes et al., 1974), with many ages ≤500 Ma. Norman et al. (2012) suggested that these locally derived, spherical glasses were produced by small impacts during an increase in the local impact flux rather than an increase in the global impact flux.

Here we present new measurements and improved interpretations of $^{40}$Ar/$^{39}$Ar ages on almost 100 lunar impact glass samples from the Apollo 14, 16, and 17 landing sites using conservative yet rigorous approaches to better understand how argon diffusion in lunar impact glass samples affects sample age. We also describe sample selection and analysis methodologies involving composition, size, and shape of lunar impact glasses. The methods described here will allow investigators to choose lunar impact glasses that are most likely to yield reliable (rather than apparent) $^{40}$Ar/$^{39}$Ar ages so that a true representation of the flux of impactors in the Earth-Moon system is revealed. Interpretations of the resultant improved flux of impactors are offered.

## 2. Selection and Characterization of Lunar Impact Glasses
### 2.1 Sample Selection
Clean, single phase glasses (not agglutinates) are prime samples for $^{40}$Ar/$^{39}$Ar analyses that investigate the lunar impact rate over time because they were heated to hyperliquidus temperatures during the melting event, were likely to have been totally degassed during that event, and were quenched to glass. When analysed, the glass contains a maximum of three Ar-isotopic components: solar wind, cosmogenic nuclides, and radiogenic $^{40}$Ar. The lunar impact glasses that we have analysed previously (Zellner et al., 2009a,b; Delano et al., 2007; Zellner et al., 2002) and in the current study (i) are not crystalline in nature (not devitrified), (ii) contain neither unmelted mineral grains (xenocrysts) nor clasts (xenoliths), (iii) do not possess crusty/dusty outer rims, and (iv) are demonstrably of impact origin (not volcanic; Delano, 1986). Geochemical data for the entire set of these ~100 samples (both analysed previously and in the current study) can be found in Appendix A. We propose in the following section, that while the selection criteria mentioned above are necessary for $^{40}$Ar/$^{39}$Ar investigations of lunar glasses, they are not sufficient. Since the extent of diffusive loss of radiogenic $^{40}$Ar from lunar glasses during residence in the near-surface of the Moon due to the duration and magnitude of diurnal temperature cycles (Figure 2) is related to the chemical composition and size of the glass (Figure 3), both of which are discussed in Section 4, it too must be considered.

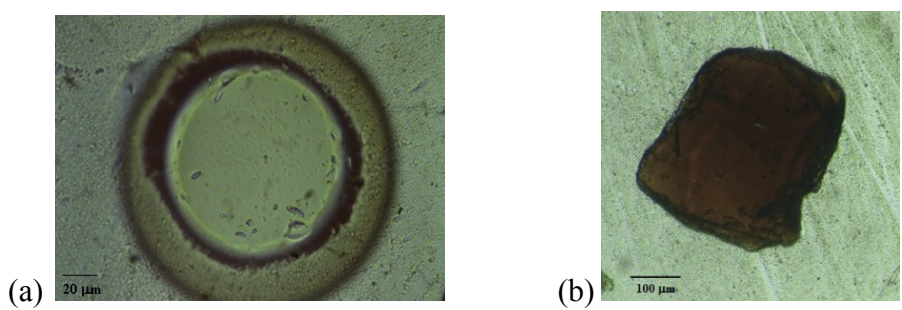

**Figure 1.** Transmitted light photomicrographs of lunar impact glasses from the Apollo 16 regolith. Note that these glasses are free of crystalline inclusions. The light green sphere (a) is 160 μm across and is compositionally similar to the local Apollo 16 regolith, while the brown glass shard (b) is 324 μm across

and has a high-Ti mare composition that is exotic to the Apollo 16 site (Delano, 1975; Zeigler et al., 2003). Both glasses are shown mounted in Crystalbond© adhesive. The sphere (a) shows the polished surface for determining the chemical composition by electron microprobe. The dark inner ring is the boundary between the polished surface of the glass and the adhesive with the glass sphere below. According to the minimum required size discussed in the text for CRE age of 750 Ma and time-averaged temperature of 290K, the glass sphere (a) with X(NBO) = 0.17 would be too small, whereas the glass shard (b) with X(NBO) = 0.33 would exceed the minimum size (Figure 3).

### 2.2 Sample Preparation

Impact glasses that have been selected using the criteria listed in the previous section were individually mounted within a sample container with CrystalBond© adhesive. Each glass was ground and polished to expose a small portion of the glass for microbeam chemical analysis. Since it is essential to maximally preserve the sample for isotopic analysis, we generally expose a polished surface of ≤50μm (Figure 1a). A planar glass surface is essential for electron microprobe analyses to determine the chemical composition of the glass. A photomicrograph of each glass provides a record of the sample that is often helpful during later stages of analysis and during preparation of the manuscript.

### 2.3 Chemical Analyses

We have used a JEOL 733 electron microprobe (Department of Earth and Environmental Sciences at Rensselaer Polytechnic Institute, Troy, NY) to determine the major-element compositions (Appendix A) of all lunar glasses that we have isotopically dated. The operating conditions have been the following: beam current = 20 nanoamps; beam diameter = 20 μm; and count-time per element using five wavelength dispersive spectrometers = 60 seconds, including peak and backgrounds for each element. Each measurement has an uncertainty of ~3% of the amount present in each sample. The time that the sample is exposed to the electron beam was ~5 minutes. In an effort to constrain the source regions of the impact glasses, it is useful to show the ratios of major elements, such as $MgO/Al_2O_3$ vs. $CaO/Al_2O_3$ (e.g., Zeigler et al., 2006; Delano 1986) or $K_2O$ (as a proxy for Th; e.g., Korotev 1998) vs. a refractory element (e.g., Zellner et al., 2009b). In addition to helping to establish relationships among glasses that may or may not be paired, determining chemical composition of glasses is essential for distinguishing volcanic glasses having picritic compositions from impact glasses that often have basaltic, noritic, and feldspathic compositions (Delano, 1986).

### 2.4 $^{40}Ar/^{39}Ar$ Ages

All of the Apollo 14, 16, and 17 lunar impact glass samples (analyzed previously by Delano et al. [2007] and Zellner et al. [2009a,b] along with those whose data are reported for the first time herein) were irradiated for ~300 hours in the Phoenix Ford Reactor at the University of Michigan; the J factors for the irradiation of these glasses were 0.05776 ± 0.00030 and 0.07857 ± 0.00048, in two separate irradiations (2002 and 2003). A small fraction of these samples was irradiated for just 80 hours in the same reactor; the J factors for this irradiation were 0.019875 ± 0.0000363, 0.0197070 ± 0.0000604, and 0.019644 ± 0.0000411, depending on the sample's location in the irradiation disk. Included along with the samples was MMhb-1 hornblende (~520 Ma; but see Jourdan and Renne [2007] for concerns about using this as a monitor) to determine the neutron fluence in the reactor, $CaF_2$ salt to correct for reactor-produced interferences, and $K_2SO_4$ to measure K interferences in the reactor. The isotopic composition of the released Ar in

6each sample was measured with a VG5400 mass spectrometer at the University of Arizona – Tucson. Each sample was degassed in a series of temperature extractions until $^{40}$Ar counts from the sample peaked and then decreased to background levels (Appendices C and D). As described in Delano et al. (2007) and Zellner et al. (2009a,b), data corrections included system blanks, radioactive decay, reactor-induced interferences, solar wind, and cosmic-ray spallation. Several spherules of Apollo 15 volcanic green glass from 15426 (e.g., Delano 1979; Steele et al. 1992), with a well-defined $^{40}$Ar/$^{39}$Ar age of ~3340 Ma (Podosek and Huneke 1973; Huneke et al. 1974; K ~200 ppm) were used as isotopic working standards. Data were reduced using Isotopic Analysis with Correlated Errors (ISAC; Hudson, 1981) and Deino software (Weirich, 2011; Deino, 2001); the decay constant of Steiger and Jäger (1977) and Renne et al. (2010) were used in the data reduction (Table 1, Appendix B).

Ages for the lunar impact glasses described herein are reported as plateau (age derived from three or more consecutive steps), weighted (average age weighted by the amount of $^{39}$Ar in each step), or one step. The uncertainties in these ages were calculated as weighted averages based on the amount of $^{39}$Ar released at each step and are reported as at least 2σ. Quality assessment (and the basis for it) for each argon release pattern is described in Section 4.2. Ages for other glasses (Hui, 2011; Ryder et al., 1996) are reported as stated in those studies. These data can be found in Table 1 and Appendices B and C.

### 2.5 Data Set
In this study, we report on the results of chemical (Appendix A) and isotopic analyses (Table 1, Appendix B) of ~100 high-K lunar impact glasses from the Apollo 14, 16, and 17 landing sites that were analyzed by our group, including measurements for 73 new ones. Data from other studies (e.g., Hui, 2011; Levine et al., 2005; Culler et al., 2000; Ryder et al., 1996) were also considered when sizes, shapes, chemical compositions, and ages, as described in Section 2.1, were known.

## 3. Formation of Lunar Impact Glasses
### 3.1 Source Material
The Apollo Soil Surveys (e.g., Reid et al., 1972a,b) reported the chemical compositions of lunar glasses extracted from lunar regoliths collected at the Apollo landing sites. Lunar glass spheres of impact origin range in size from ≤25 μm (Keller and McKay, 1992) to ~6 mm (Ryder et al., 1996). However, it is not known in what size impact or from what kind of material the glasses are produced. Compositional clusters of glasses, usually of impact origin, were interpreted as reflecting the compositions of *rocks* in the target (e.g., Reid et al., 1972a,b). In contrast, other investigators (e.g., Korotev et al., 2010; Zellner et al., 2009a; Delano et al., 2007; Zeigler et al., 2006; Zellner et al., 2002) have observed that impact-generated glasses commonly have chemical compositions similar to that of the local *regoliths*, not necessarily of one or a few individual rocks. In addition to weakening the claim by Hörz and Cintala (1997) that there is a paucity of glasses having regolith compositions, that observation is consistent with theoretical modelling (Wünnemann et al., 2008) showing that porous target-materials (e.g., lunar regoliths) generate higher melt volumes than non-porous targets at a given impact energy. Lunar regoliths in the uppermost 3-meters of the Moon have porosities ~37% (e.g., Mitchell et al., 1972) and densities ~1.8-2.0 g/cm$^3$ (e.g., Mitchell et al., 1972).

## 3.2 Crater Size

The sizes of the craters that produce lunar impact glasses are unknown but they can provide insight into the size of the impactor that created each glass and the resultant shape of the glass. One thought is that impact glasses are formed only in cratering events <1 km in diameter (e.g., Norman et al., 2012; Hörz and Cintala, 1997). Micrometeorite impacts, in particular, however, seem unlikely to generate significantly large volumes of lunar impact glasses (e.g., ~$3\times10^7$ μm$^3$ for a 400-μm diameter glass spherule), the type of which are described here. Other investigators prefer a range of crater sizes (<1 m to >100 km), especially if the glass composition is clearly exotic to the local regolith in which it was found (e.g., Korotev et al., 2010; Delano et al., 2007; Zeigler et al., 2006; Delano, 1991; Symes et al., 1988). Korotev et al. (2010) found that ~75% of the impact glass in the Apollo 16 regolith is compositionally different from any mixture of rocks from which the regolith is mainly composed. Therefore, those impact glasses have been interpreted as being exotic to the Apollo 16 region and probably were formed by, and ballistically transported from, cratering events ≥100 km from the landing site (Korotev et al., 2010; Delano et al., 2007; Zeigler et al., 2006); Delano et al. (2007) found the majority of those exotic glasses to be non-spherical (i.e., shards). The shapes of the glasses reported herein have been used to suggest source terrain(s) as well as likelihood to report true $^{40}$Ar/$^{39}$Ar ages.

# 4. Results

## 4.1 Chemical composition and size: Implications for interpreting $^{40}$Ar/$^{39}$Ar ages in lunar impact glasses

All previous investigators (e.g., Zellner et al., 2009a,b; Delano et al., 2007; Levine et al., 2005; Culler et al., 2000) have implicitly assumed that lunar impact glasses are highly retentive of radiogenic $^{40}$Ar during prolonged residence in the shallow lunar regolith that is subjected to diurnal temperature variations. However, the rate of Ar diffusion was experimentally measured by Gombosi et al. (2015) in three large (~1.6 mm diameter), inclusion-free, lunar impact glass spherules having uniform chemical compositions similar to that of the average Apollo 16 regolith with and X(NBO) value of ~0.18. That investigation showed that significant loss of radiogenic $^{40}$Ar would occur during some exposure histories, such as ~75% loss from a 400-μm diameter glass spherule residing at <2-cm depth below the lunar surface for 40 Ma. Figure 2 shows the range of diurnal temperature variations in the lunar regolith near the Moon's equator. The magnitude of the temperature variations diminishes with depth to a nearly constant temperature of 260K (-15°C) at a depth of ~60 cm (Vasavada et al., 2012; Lawson et al., 1999; Langseth et al., 1976).

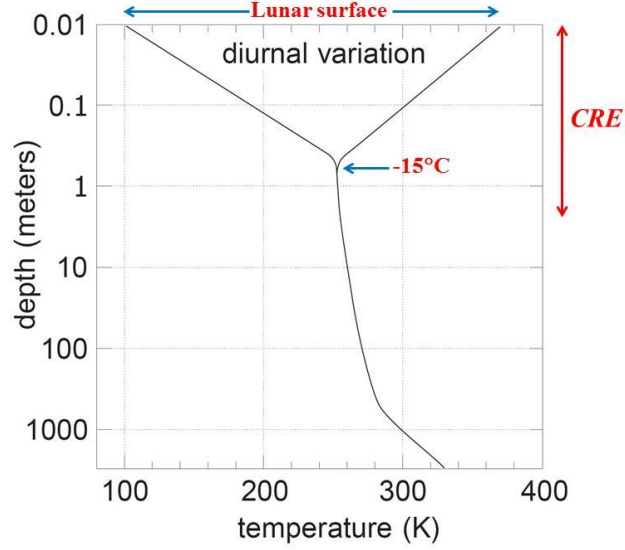

**Figure 2.** Diurnal temperature cycles near the Moon's equator occur in the upper ~60 cm of the lunar regolith. The magnitude of the temperature cycle diminishes with depth to a nearly constant temperature of 260 K (-15°C) at a depth of ~80 cm (Vasavada et al., 2012; Lawson et al., 1999; Langseth et al., 1976). The absolute temperatures and magnitude of diurnal cycling decrease with increasing latitude. Cosmic ray exposure (CRE) age is the time that a sample has resided within the upper few meters of the regolith (Eugster, 2003).

Diffusivity of radiogenic $^{40}$Ar depends on chemical composition and melt structure, which can be parameterized using the fraction of non-bridging oxygens, X(NBO) (Lee, 2011; Mysen and Richet, 2005). As shown in Eq. 1, for a given temperature, the Ar diffusivity of a glass is inversely proportional to its X(NBO) value (Lee, 2011):

$$X(NBO) = \frac{2 \times (X_{NC} - X_{FC})}{2 - (X_{NC} - X_{FC})} \qquad [\text{Eq. 1}]$$

where $X_{NC}$ = mole fraction of oxide with cations having network-modifying and charge-balancing roles (e.g., FeO, MnO, MgO, CaO, Na$_2$O, and K$_2$O); and $X_{FC}$ = mole fraction of oxide with cations having network-forming roles other than Si (e.g., TiO$_2$ and Al$_2$O$_3$; Lee, 2011 and references therein). Since Cr$^{2+}$ is known to be the dominant valence state of Cr in lunar materials (e.g., Sutton et al., 1993; Smith, 1974), CrO was included as an additional component, albeit a minor one, in the $X_{NC}$ term. Titanium, which can be abundant in some lunar materials, was assumed to contribute entirely to the $X_{FC}$ component (e.g., Farges et al., 1996).

To estimate the temperature-,time-integrated, Ar-diffusion coefficient of lunar glasses as a function of X(NBO), it was assumed that the main process for causing Ar loss in lunar glasses was thermal diffusion of Ar during the CRE (cosmic ray exposure) in the shallow lunar regolith, rather than episodic shock events. For lunar glass spheres with uniform abundances of K, the fraction of total $^{40}$Ar lost, **f**, during that residence in the shallow lunar regolith was determined by step-heating of the glass spheres. The equation (McDougall and Harrison, 1999) used to estimate the temperature-,time-integrated Ar diffusion coefficient, **D**, for lunar glass spheres (e.g., Huneke, 1978) with known radii, CRE ages, fraction of Ar lost, and X(NBO) is shown below:

$$D = \frac{a^2}{\pi^2 t}\left(2\pi - \frac{\pi^2}{3}f - 2\pi\sqrt{1-\frac{\pi}{3}f}\right) \quad \text{for } f \leq 0.85 \quad \textbf{[Eq. 2]}$$

Here **a** = radius (cm) of the glass sphere, **t** = time (seconds) spent in the shallow lunar regolith when diffusive Ar loss occurred as recorded by the CRE age of the glass, and **f** = fraction of Ar lost during the glass' post-formation thermal history (e.g., burial in lunar regolith). If the CRE age of a lunar sample has been calculated based on spallation production rates at the lunar surface, then the actual time the sample spent within ~1-2 meters of the lunar surface (Figure 2) would be greater (Podosek and Huneke, 1973) and the calculated **D** would be an upper limit (since **D** is inversely proportional to time and decreases at T decreases).

The results are shown in Figure 3, where temperature contours appropriate for the uppermost ~60 cm of lunar regolith (Figure 2) are shown. While the rate of Ar diffusion with temperature is known for lunar glasses having X(NBO) ~0.18-0.19 (Gombosi et al., 2015), its dependence over the observed range of X(NBO) for lunar glasses has been inferred using the trend defined by the calculated temperature-,time-integrated Ar diffusivity, represented by log D(T,t), of several lunar glasses, as described below. The absolute temperatures associated with each contour are based on the results from Gombosi et al. (2015). The lunar glasses had diameters ranging from 80 μm to >1400 μm and CRE ages ranging from 30 Ma to 300 Ma.

The two main goals of Figure 3 are to (i) estimate the diffusivity of $^{40}$Ar in lunar glasses as a function of chemical composition, X(NBO), and to (ii) use that information to guide the selection of lunar glasses for $^{40}$Ar/$^{39}$Ar dating in order to find those that have experienced minimal loss of $^{40}$Ar. The strategy for this estimation is based on using lunar glasses of known dimensions, CRE age, fraction of $^{40}$Ar lost, chemical composition, and shape (sphere or shard) to estimate the temperature-,time-integrated Ar diffusion coefficient, represented here by log D(T,t). In generating the model illustrated in Figure 3, it was assumed that diffusive loss of $^{40}$Ar from the glasses occurred as a result of their having resided within the thermal regime of the upper 1-2 meters of the lunar regolith for a time recorded by their CRE ages, i.e., **t** in the diffusion equation (Eq. 2 above).

The samples plotted in Figure 3 are described in the following paragraphs. With additional Ar-isotopic data on actual lunar glasses and additional experimental work on Ar diffusion in lunar glasses (and compositional analogues), especially at high values of X(NBO) ~ 0.50-0.60, the slope of the isotherms will become better constrained.

**4.1.1 Apollo 16 Impact Glass (61502,13,3)**
Ar diffusion in this glass sphere (chemically homogeneous and clast-free) with radius ~735 μm was reported by Gombosi et al. (2015). The chemical composition is similar to that of the local Apollo 16 regolith with X(NBO) = 0.187 (solid circles in Figure 3). The values for log D(T,t) for this glass as a function of temperature, which were calculated using the experimental results of Gombosi et al. (2015), are shown by the nine points at X(NBO) = 0.187 in Figure 3. Those points tightly constrain Ar diffusivity at the low end of the X(NBO) range observed in lunar glasses.

### 4.1.2 Apollo 15 Volcanic Green Glass (15426)
Spheres (chemically homogeneous and clast-free) of this low-Ti picritic glass (e.g., Delano, 1979; Steele et al., 1992) have an $^{40}Ar/^{39}Ar$ age of 3.38 ± 0.06 Gy (Podosek and Huneke, 1973) and a CRE age ~300 My (Podosek and Huneke, 1973; Spangler et al., 1984). The dominant compositional group ('A' of Delano, 1979) among this suite of picritic volcanic glasses has X(NBO) = 0.598 (open star in Figure 3). Podosek and Huneke (1973) analyzed green glass spheres with diameters ranging from 250 µm to 750 µm, and used 400 µm for much of their discussion. Using a radius = 200 µm, CRE age = 300 My, and fraction of $^{40}Ar$ lost = 0.02 ± 0.01 (Podosek and Huneke, 1973), the log D(T,t) = -23.5 to -24.4. With this range, Figure 3 shows that green glass spheres with diameters of at least 65-185 µm would have lost ≤10% of their $^{40}Ar$ in 750 My with that range of log D(T,t).

### 4.1.3 Apollo 17 Volcanic Orange Glass (74220)
Spheres (chemically homogeneous and clast-free) of this high-Ti picritic glass (e.g., Delano, 1986; Heiken et al., 1974) have an $^{40}Ar/^{39}Ar$ age of 3.60 ± 0.04 Gy (Huneke, 1978) and a CRE age ~30 My (Huneke, 1978; Eugster et al., 1979). Using a sphere with radius = 40 µm based on the mass of individual glasses analyzed by Huneke (1978), X(NBO) = 0.505 (Delano, 1986), CRE age = 30 My, and estimated fraction of $^{40}Ar$ lost ~0.03–0.07, the value of log D(T,t) = -23.1 to -23.9 (open square in Figure 3). With this range of log D(T,t) values, Figure 3 shows that orange glass spheres with diameters of at least 120 µm – 280 µm would have lost ≤10% of their $^{40}Ar$ in 750 My with that range of log D(T,t).

### 4.1.4 Apollo 17 Impact Glass, C6/301 (71501)
This sphere (chemically homogeneous and clast-free) is a light green glass with X(NBO) = 0.248, CRE age = 75 ± 10 My, $^{40}Ar/^{39}Ar$ age = 102 ± 20 My, diameter = 360 µm, and fraction of $^{40}Ar$ lost = 0.24. These characteristics yielded log D(T,t) = -21.0 to -21.2 (solid star in Figure 3), showing that a glass with this composition would require a minimum diameter of 2700 µm – 3100 µm to have lost ≤10% of its $^{40}Ar$ in 750 My with that range of log D(T,t).

### 4.1.5 Apollo 16 Impact Glass, G3/225 (64501)
This angular shard (chemically homogeneous and clast-free), a brown glass with X(NBO) = 0.201, CRE age = 145 ± 20 My, $^{40}Ar/^{39}Ar$ age = 3739 ± 20 My, and average dimension = 184 µm, was reported by Delano et al. (2007). This glass (open triangle in Figure 3), which belongs to a distinctive suite of impact glasses at the Apollo 16 site (Delano et al., 2007; Zeigler et al., 2006), had lost ≤1% of its $^{40}Ar$. These characteristics yielded log D(T,t) = -24.7 to -25.4. Figure 3 shows that a glass with this composition would require dimensions of only 20 µm – 40 µm to have lost ≤10% of its $^{40}Ar$ in 750 My with that range of log D(T,t). The implication for this glass is that it had been spent most of its CRE history at low temperatures insulated from diurnal temperature variations by the overlying regolith.

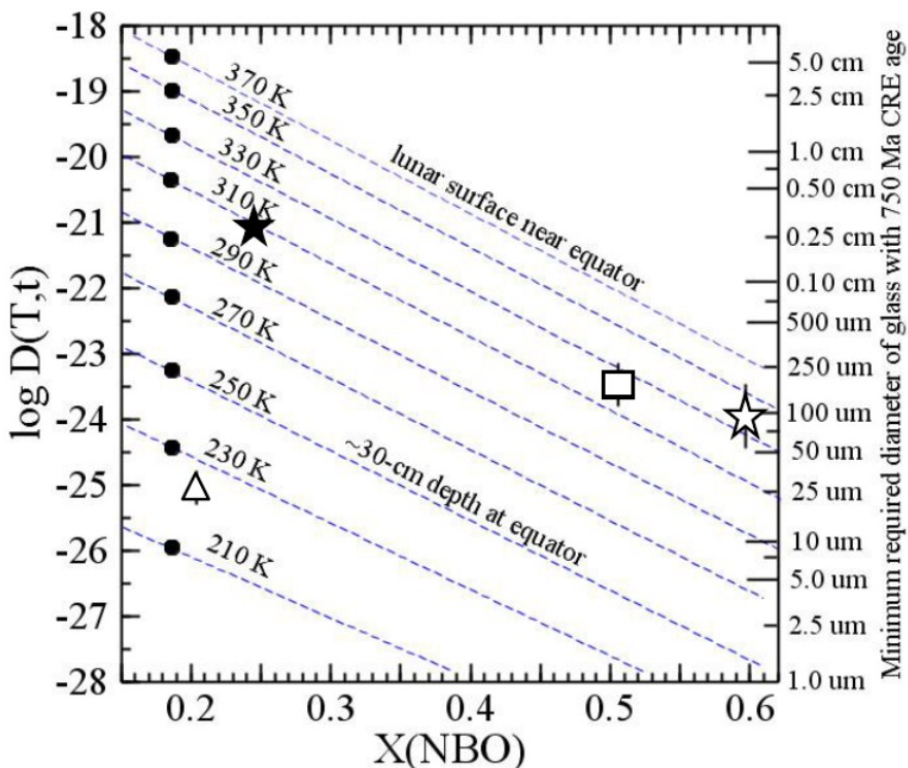

**Figure 3.** Values for the temperature-,time-integrated Ar diffusion coefficient, log D(T,t), in lunar glasses have been determined using their measured diameters, chemical compositions, CRE ages, and % Ar lost by thermal diffusion during their residence time in the shallow lunar regolith. The lunar glasses encompass a large range of (i) chemical composition, (ii) CRE ages, and (iii) $^{40}$Ar/$^{39}$Ar ages. The results show a strong compositional dependence on log D(T,t) using the fraction of non-bridging oxygens, X(NBO). The minimum sizes of glasses required to retain at least 90% of their radiogenic $^{40}$Ar during CRE ages of 750 Ma for a range of temperatures and compositions are shown on the right side. All of the glasses have dimensions far in excess of the minimum sizes required for their compositions and CRE ages. As described in the text, the solid circles represent an Apollo 16 impact glass (61502,13,3); the open star represents an Apollo 15 volcanic green glass (15426); the open square represents an Apollo 17 volcanic orange glass (74220); the solid star represents an Apollo 17 impact glass (C6/301, 71501); and the open triangle represents an Apollo 16 impact glass (G3/225, 64501). Uncertainties on log D(T,t) for the lunar glasses, which are controlled by uncertainties in the CRE ages, are similar to the height of the symbols. The lunar volcanic glasses are not plotted in the subsequent figures involving lunar impact glasses exclusively.

## 4.2 Interpreting $^{40}$Ar/$^{39}$Ar Data

We do not know whether the data for the lunar glasses shown in Figure 3 are typical for the regolith-gardening process since the end of the late heavy bombardment. However, the slope of the isotherms (Figure 3) suggests that the Apollo 15 green volcanic glass, Apollo 17 orange volcanic glass, and impact glass sphere C6/301 all resided at comparably shallow depths in the lunar regolith during their temperature-,time-integrated CRE histories. Those three glasses retained >75% of their radiogenic $^{40}$Ar to yield reliable ages. Glass shard G3/225 resided at a greater depth (i.e., cooler) in the lunar regolith that allowed this glass to retain ≥99% of its $^{40}$Ar, and a reliable $^{40}$Ar/$^{39}$Ar age.

With this model of argon diffusivity as a guide, the current investigation revisits $^{40}$Ar/$^{39}$Ar ages on 22 lunar impact glasses (Delano et al. 2007; Zellner et al. 2009a,b) and introduces ages for 73 new ones from the Apollo 14, 16, and 17 landing sites (Table 1, Appendices B and C). These 98 glasses were not only free of exotic components, such as unmelted crystals and lithic fragments derived from the impacted target, but also had known sizes, shapes, and chemical compositions (Section 2.1). After laser step-heating on these 98 impact glasses, 85 yielded $^{40}$Ar/$^{39}$Ar ages (single-step, plateau, or weighted), 10 yielded indeterminate "young" ages, and three yielded no ages. In an effort to distinguish those impact glasses that have a stronger likelihood of having retained a reliable $^{40}$Ar/$^{39}$Ar age of impact formation from those that did not, the minimum size associated with an exposure scenario of a 750-Ma CRE history (Figure 3) has been applied as a selection criterion to those 95 impact glasses that yielded ages, as well as to impact glasses from other studies (e.g., Hui, 2011; Ryder et al., 1996). Evaluative assessments for each age determination are given in Table 1 and Appendix B, where argon release patterns were deemed "good" if >50% $^{39}$Ar was used in the age and most of the steps were concordant; "fair" if some of the steps were concordant; and "poor" if none of the steps were concordant. Only ages determined to be "good" or "fair" are included in the following figures, except where small size excludes the sample. Figures 4 and 5 illustrate which of these impact glasses were large enough to have retained at least 90% of their radiogenic $^{40}$Ar during that model exposure history, and which ones were not of sufficient size. As noted in Delano et al. (2007), the shapes of the lunar impact glasses have been described as being either spherical (Figure 4) or broken shards (Figure 5). Among the glass spheres (Figure 4), only ~40% are likely to have accurately recorded their ages of impact formation.

Figure 4 shows that most of the impact glass spheres that did not satisfy the minimum required size to have retained at least 90% of their radiogenic $^{40}$Ar during a 750-Ma exposure age have chemical compositions with X(NBO) <0.25 (open symbols in Figure 4). Those lunar glasses have lunar highlands feldspathic compositions with higher Ar diffusivities at a given temperature than more mafic glasses with higher X(NBO) values and lower Ar diffusivities (Figure 3). When the minimum size criterion for the same exposure scenario was applied to the impact glass shards (Figure 5), ~60% of those analyzed impact glasses were found to satisfy the minimum required size criterion. As expected, most of the impact glass shards that were found to be too small and were likely to have lost $^{40}$Ar* had X(NBO) <0.25 (Figure 5).

Figure 6 is a compilation of the impact glass spheres (Figure 4) and shards (Figure 5) that exceeded the minimum size requirement for the model exposure history. Consequently, the impact glasses in Figure 6 are considered likely to have yielded reliable $^{40}$Ar/$^{39}$Ar ages. Figure 7 shows a histogram of the resulting age-frequency distribution of those impact glasses that satisfied the minimum required size criterion.

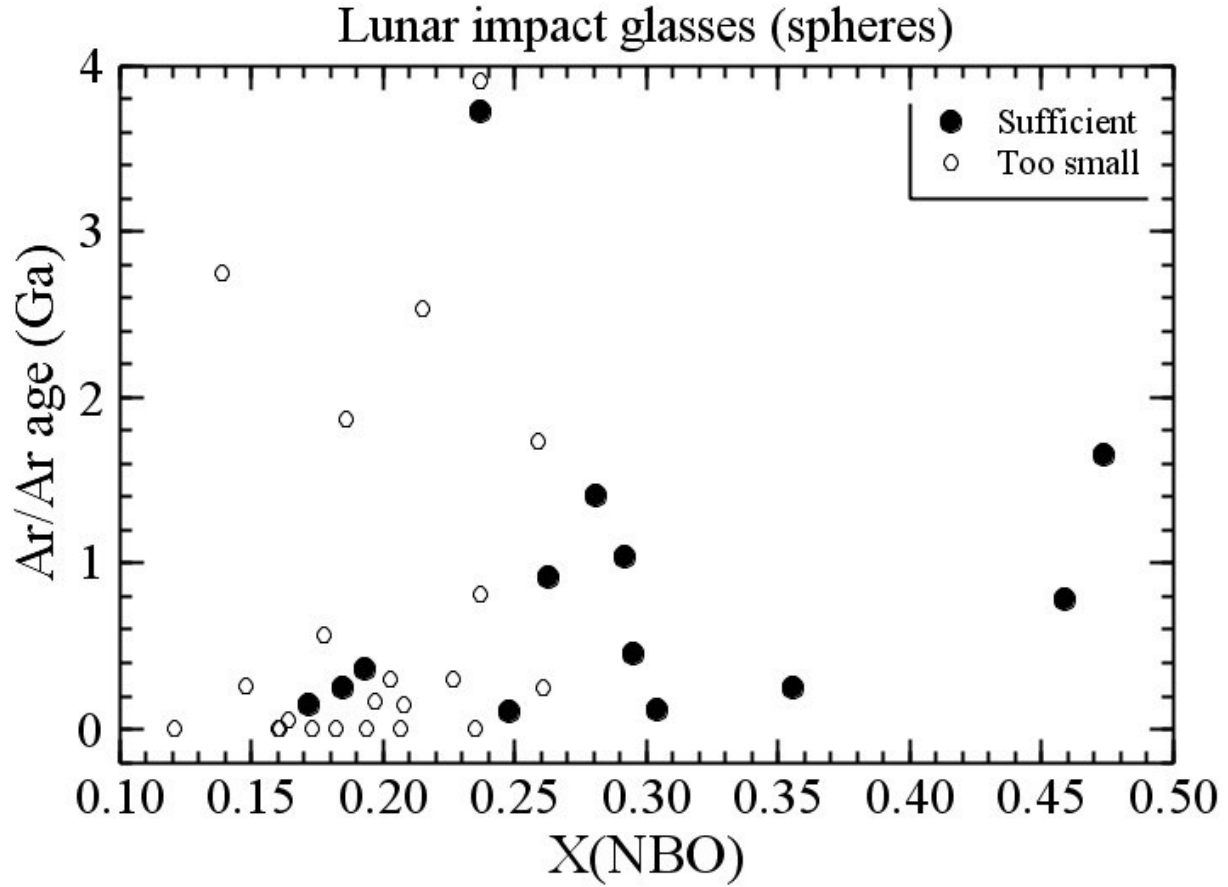

**Figure 4.** Lunar glass *spheres* that have been analyzed by Zellner et al. (2009a, 2009b), Hiu (2011), and Ryder et al. (1996) with known chemical compositions, dimensions, and $^{40}$Ar/$^{39}$Ar ages have been plotted, along with spheres from this study (Table 1, Appendices A and B). Glass spheres having sufficient sizes that could have retained at least 90% of their radiogenic $^{40}$Ar following 750 Ma in the shallow lunar regolith at a time-integrated temperature of up to 290K (Figure 3) are indicated by solid symbols. Glass spheres that would have been too small to have retained at least 90% of their radiogenic $^{40}$Ar during that temperature, time history are shown by open symbols.

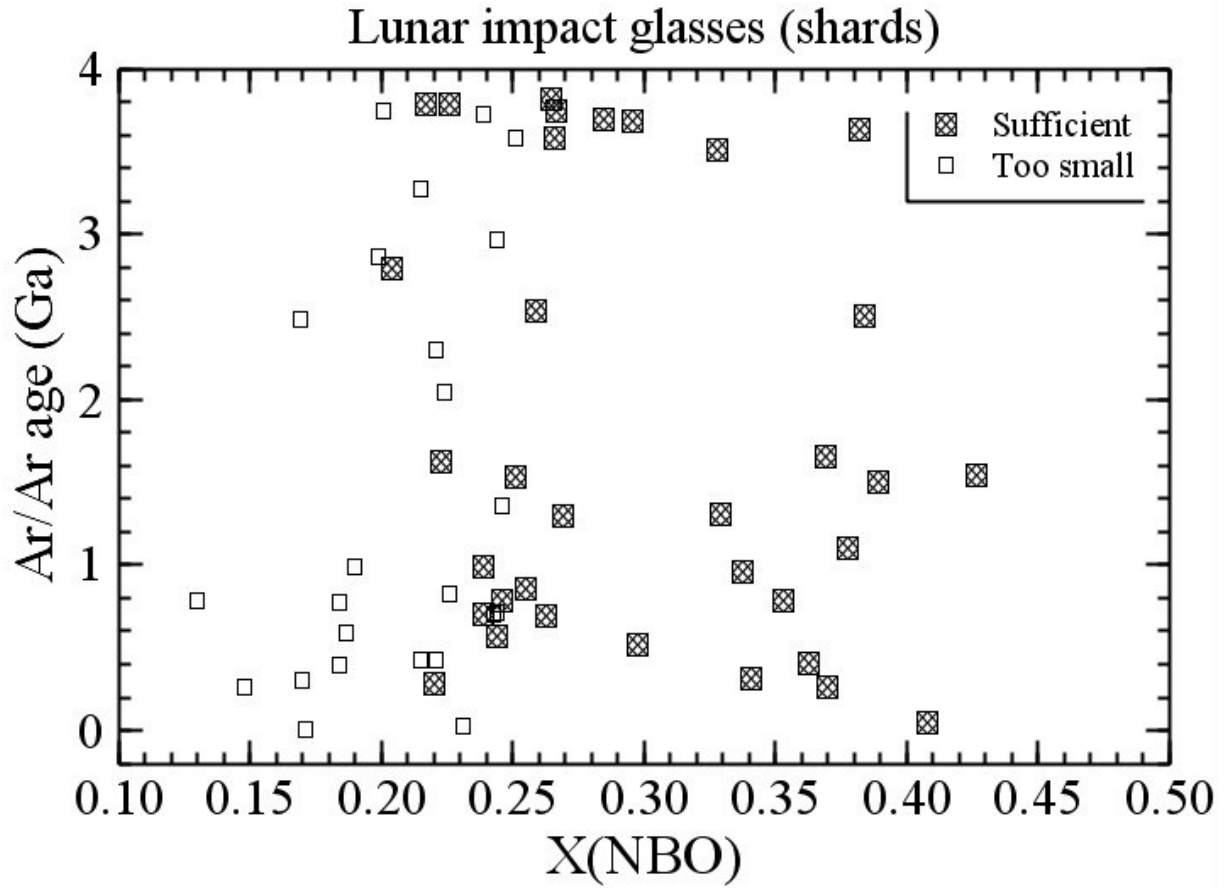

**Figure 5.** Lunar glass *shards* that have been analyzed by Zellner et al. (2009a, 2009b) and Delano et al. (2007) with known chemical compositions, dimensions, and $^{40}Ar/^{39}Ar$ ages have been plotted, along with shards from this study (Table 1, Appendices A and B). Glass shards having sufficient sizes that could have retained at least 90% of their radiogenic $^{40}Ar$ following 750 Ma in the shallow lunar regolith at a time-integrated temperature of up to 290K (Figure 3) are indicated by partially filled boxes. Glass shards that would have been too small to have retained at least 90% of their radiogenic $^{40}Ar$ during that temperature, time history are shown by open symbols.

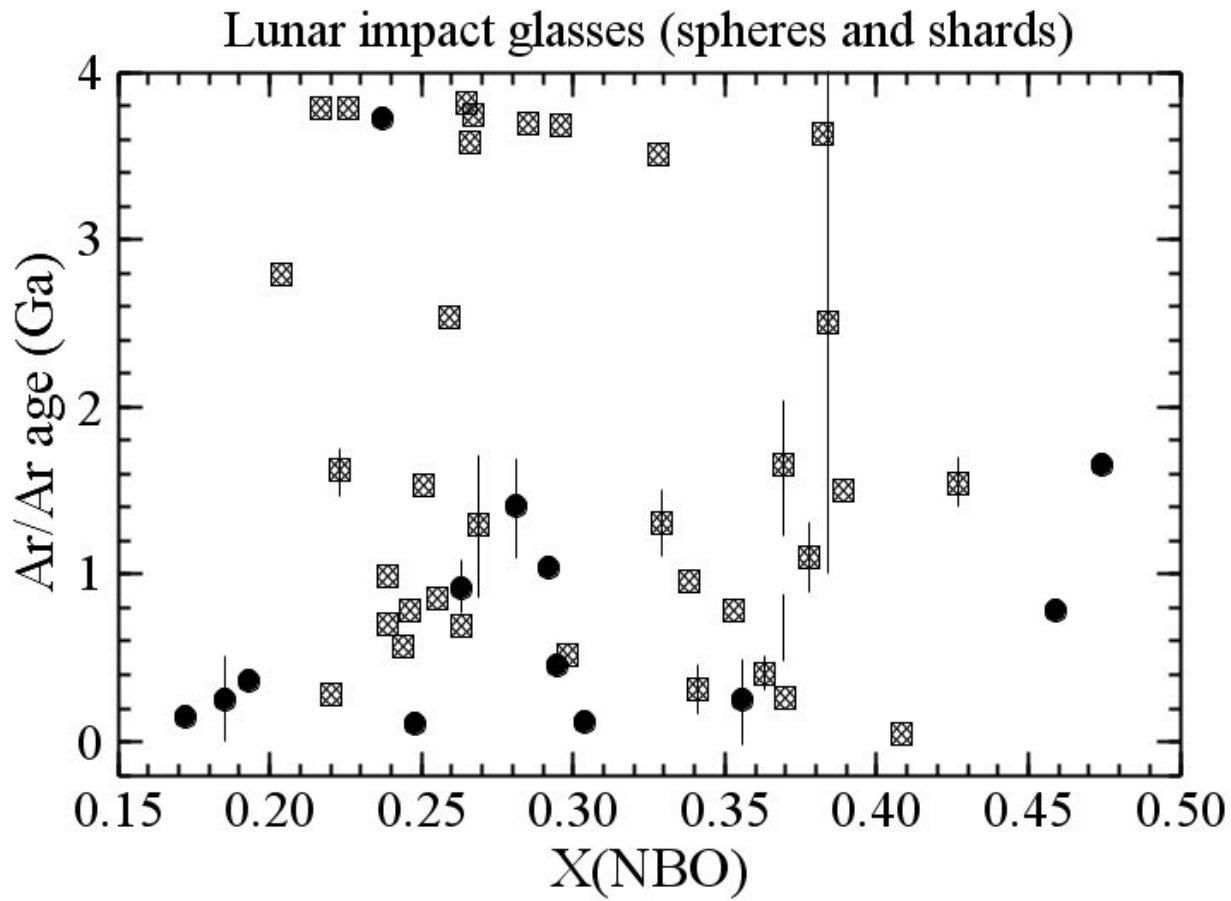

**Figure 6.** Compilation of lunar impact glass spheres (solid circles; see Figure 4) and lunar impact glass shards (partially filled boxes; see Figure 5) that would have likely retained at least 90% of their radiogenic $^{40}$Ar during 750 Ma of residence at a time-integrated temperature of ~290K (Table 1, Appendix A). Uncertainties in age that are larger than the size of the symbols are shown.

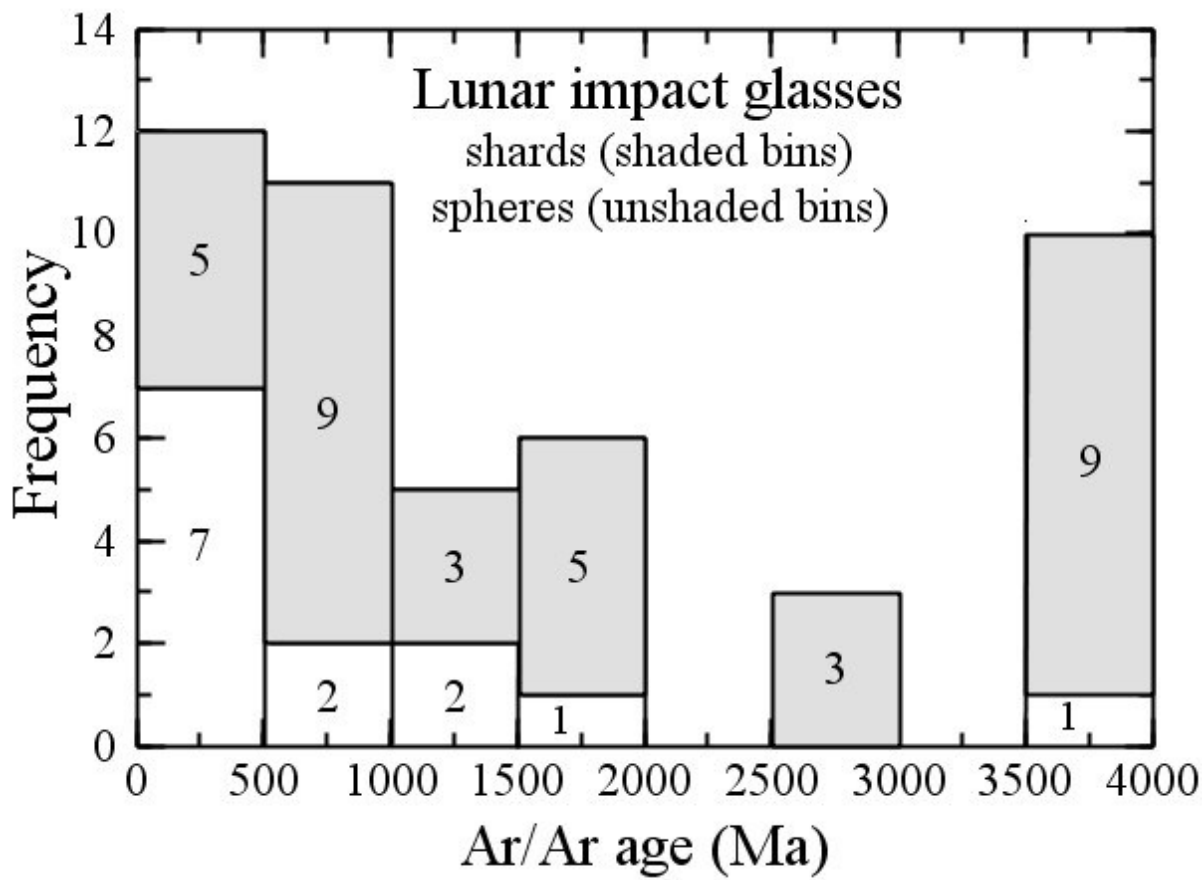

**Figure 7.** Age-frequency distribution of lunar impact glass spheres (unshaded bins) and lunar impact glass shards (shaded bins) that exceed the minimum sizes required to have retained at least 90% of their radiogenic $^{40}$Ar during 750 Ma at a time-integrated temperature of 290K (Table 1, Figure 6). These lunar impact glasses are likely to have yielded accurate ages of the impact events that generated the melts. The number of impact glasses within each bin is shown.

## 5. Discussion
### 5.1 Implications for $^{40}$Ar/$^{39}$Ar Dating of Lunar Impact Glasses

The chemical compositions not only distinguish between impact-generated glasses and volcanic glasses (an essential distinction if impact flux is the focus of an investigation; Delano, 1986). By knowing X(NBO), the minimum size of glass required to yield an accurate $^{40}$Ar/$^{39}$Ar age of impact melting can be estimated (Figure 3). For example, lunar impact glasses with X(NBO) ≤ 0.25 are dominantly feldspathic highlands compositions (e.g., anorthosite-norite-troctolite; Wu et al., 2012; Taylor, 2009; Prettyman et al., 2006; Korotev, 2005) and are thus most susceptible to diffusive loss of radiogenic $^{40}$Ar during extended residence in the shallow (<2-cm depth; Gombosi et al., 2015) regolith during diurnal temperature variations (Figures 2, 3). The effect of greater diffusion for glasses with low X(NBO) values is clearly evident in Figures 4 and 5 where the majority of impact glasses with X(NBO) < 0.25 did not satisfy the minimum size criterion, and hence were likely to have yielded apparent, rather than true, $^{40}$Ar/$^{39}$Ar ages. In contrast, lunar picritic volcanic glasses with X(NBO) ~ 0.39-0.60 (e.g., Apollo 15 green A = 0.598; Apollo 15 yellow = 0.524; Apollo 17 orange = 0.505; refer to Delano, 1986 for the 25 known varieties) and diameters often <250 μm yield $^{40}Ar/^{39}Ar$ eruption ages (3300-3700 Ma; Spangler et al., 1984; Huneke, 1978; Husain and Schaeffer, 1973; Podosek and Huneke, 1973) that consistently overlap the $^{87}Rb/^{87}Sr$ and/or $^{147}Sm/^{143}Nd$ ages of the local crystalline mare basalts (Nyquist and Shih, 1992; Papanastassiou et al., 1977). This empirical observation provides strong additional evidence for the observed relationship (Figure 3) that Ar diffusivity decreases sharply with increasing X(NBO). While the minimum size of glass as a function of X(NBO) has been estimated (Figure 3) for a stringent temperature-time exposure history, additional experimental work on lunar-relevant compositions, preferably actual lunar glass spheres, is needed to better define Ar diffusivity in glass as a function of X(NBO).

**5.2 Young (<500 Ma) Lunar Impact Glass Spheres**
**5.2.1 Increased Cratering Rate vs. Thermal Strain**
When analyzing lunar impact glasses from a single landing site, the shapes (spherules vs. shards) and chemical compositions (local vs. exotic) of impact glasses become especially important criteria to consider when developing hypotheses about the global lunar impact flux over time. On the basis of $^{40}Ar/^{39}Ar$ ages of glass spherules from the Apollo 14 landing site, Culler et al. (2000) and Muller et al. (2001) concluded that the cratering flux has increased by a factor of ~3 in the last 500 Ma. $^{40}Ar/^{39}Ar$ ages of lunar impact glass spheres from the Apollo 12 (Levine et al., 2005) and Apollo 16 (Hui et al., 2010) landing sites, and U-Th-Pb "chemical ages" of lunar impact glass spheres from the Apollo 17 (Norman et al., 2012) landing site, were also interpreted as being consistent with an increased flux in the last 500 Ma. While the lunar and terrestrial cratering records have also been used as possible evidence for a factor-of-two increase in the cratering rate during the last ~500 Ma (McEwen et al., 1997; Grieve and Shoemaker, 1994), the issue remains unresolved (Bland 2005; Grier and McEwen, 2001).

Although the results of the current investigation also show a strong increase in the frequency of lunar impact glass spheres with $^{40}Ar/^{39}Ar$ ages <500 Ma (Figure 7), an alternative explanation is offered. We hypothesize that lunar impact glass spheres are intrinsically prone to breaking into shards, and hence have geologically short lifespans. Evidence in support of this notion comes from differential thermal analysis of lunar impact glasses showing that lunar impact glasses contain high thermal stresses (Ulrich, 1974; strain exotherms) caused by rapid quenching from hyperliquidus temperatures. These thermal stresses would make impact-generated glass spheres susceptible to breaking into shards. The impact glass spheres are broadly analogous to the inexpensive glassware that fractures spontaneously in the laboratory because the thermal stresses induced during the manufacturing process have not been effectively removed by subsequent annealing. Consequently, *lunar impact glass spheres would be expected to be short-lived*. If correct, the high rate of occurrence of lunar impact glass spheres with ages <500 Ma, as reported by previous workers and evident in Figure 7, need not require a substantial increase in the impact flux during the last ~500 Ma.

In contrast to the preponderance of impact-produced glass spheres with ages <500 Ma, lunar volcanic glass spheres have $^{40}Ar/^{39}Ar$ ages in the range of 3300-3700 Ma (Spangler et al., 1984; Huneke 1978; Husain and Schaeffer, 1973; Podosek and Huneke, 1973). Unlike impact-produced glass spheres, lunar volcanic glass spheres have lower strain exotherms (Ulrich, 1974) that cause those glasses to be less susceptible to spontaneously breaking into shards. This lower strain is possibly related to lunar volcanic glass spheres having been partially annealed in a warm pyroclastic deposit following their quenching from near-liquidus temperatures (Arndt et al., 1984).

**5.2.2 Effect of Minimum Size Criterion on Impact Flux Curves**
Relative age plots (referred to in some of the literature as "ideograms") have been used frequently to illustrate the impact flux as reported by lunar impact glasses, lunar meteorites, and asteroidal meteorites and can be influenced by one or two samples with well-defined ages; these samples show up as "spikes" and point misleadingly to an enhanced impact flux. Figure 8a shows a relative age plot for the ~100 lunar impact glasses reported here (Table 1, Appendix B). Multiple spikes are seen in the data, especially at younger ages.

Figure 8b, on the other hand, shows the age distribution of 48 lunar impact glass spheres and shards (Table 1, Figures 6 and 7) from the Apollo 14, 15, 16, and 17 landing sites that have satisfied our minimum required size criterion. The elimination of impact glasses that were too small and thus lost an appreciable fraction of $^{40}Ar^*$ significantly decreases the frequency of impact ages <1000 Ma (compare Figures 8a and 8b, which have 64 and 25 samples with ages <1000 Ma, respectively) while increasing the signal-to-noise ratio overall. Since most of the impact glasses found to have been most vulnerable to diffusive loss of radiogenic $^{40}Ar$ were associated with $^{40}Ar/^{39}Ar$ ages <1000 Ma (Figures 4, 5, Appendix B), it is not surprising that the relative probability for glass ages <1000 Ma is less in Figure 8b.

We propose that this relative age plot (Figure 8b) shows a plausible distribution of currently available ages among lunar impact glasses of sufficient size depending on X(NBO) value. It is different in appearance from any of the other relative age plots of lunar impact glasses that have been shown by other investigators (Norman et al., 2011; Hui 2011; Hui et al., 2010; Zellner et al., 2009b; Levine et al., 2005; Culler et al., 2000). Specifically, though there are young ages, the plot shows no indication of an obvious increase in the impact flux in the most recent 500 Ma. Peaks representing young ages can be shown (with careful comparison of age and composition) to be influenced by just one glass with a well-defined age and small uncertainty. Other peaks (red arrows; Figure 8b) represent multiple glasses with similar ages and different compositions. Figure 8b shows impact episodes that have been well documented elsewhere (e.g., ~500 Ma [Schmitz et al., 2001, 2003]; ~800 Ma [Zellner et al., 2009b; Swindle et al. 2009]; ~3700 Ma [Delano et al., 2007]) but with improved signal-to-noise ratio.

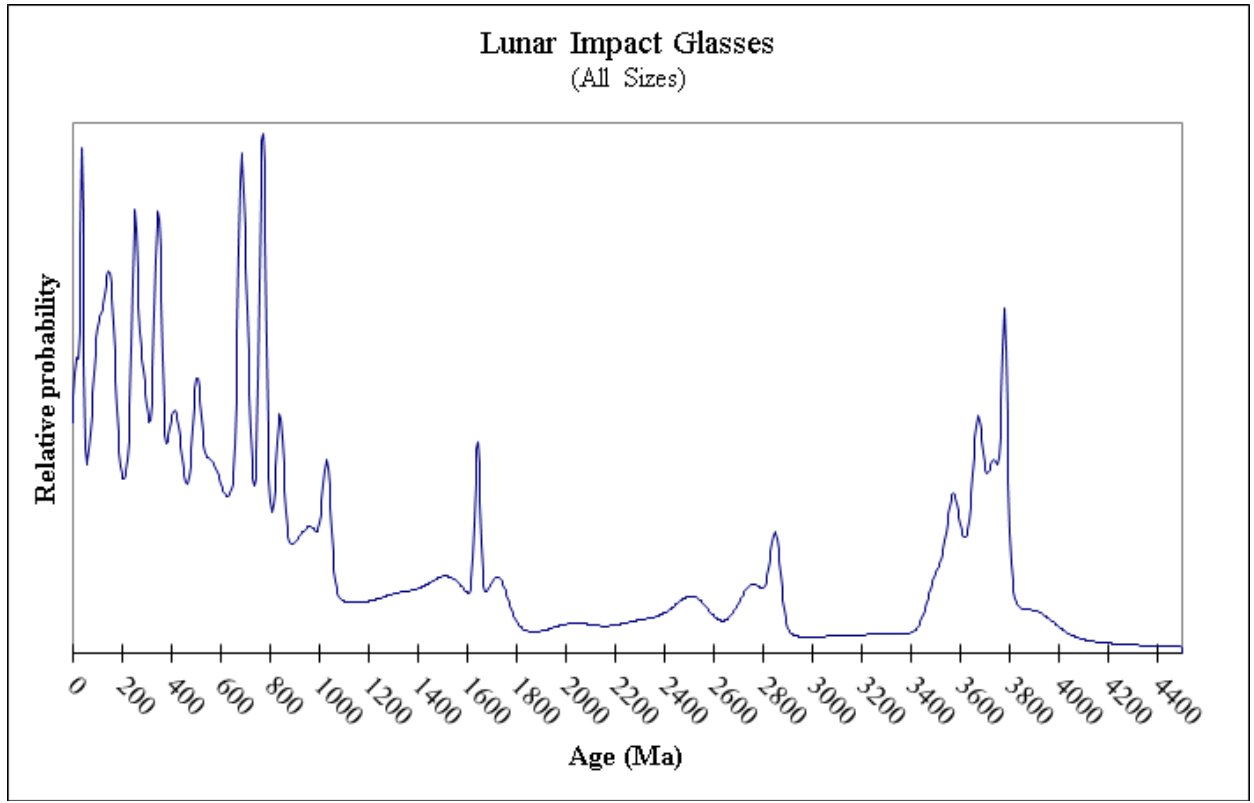

(a)

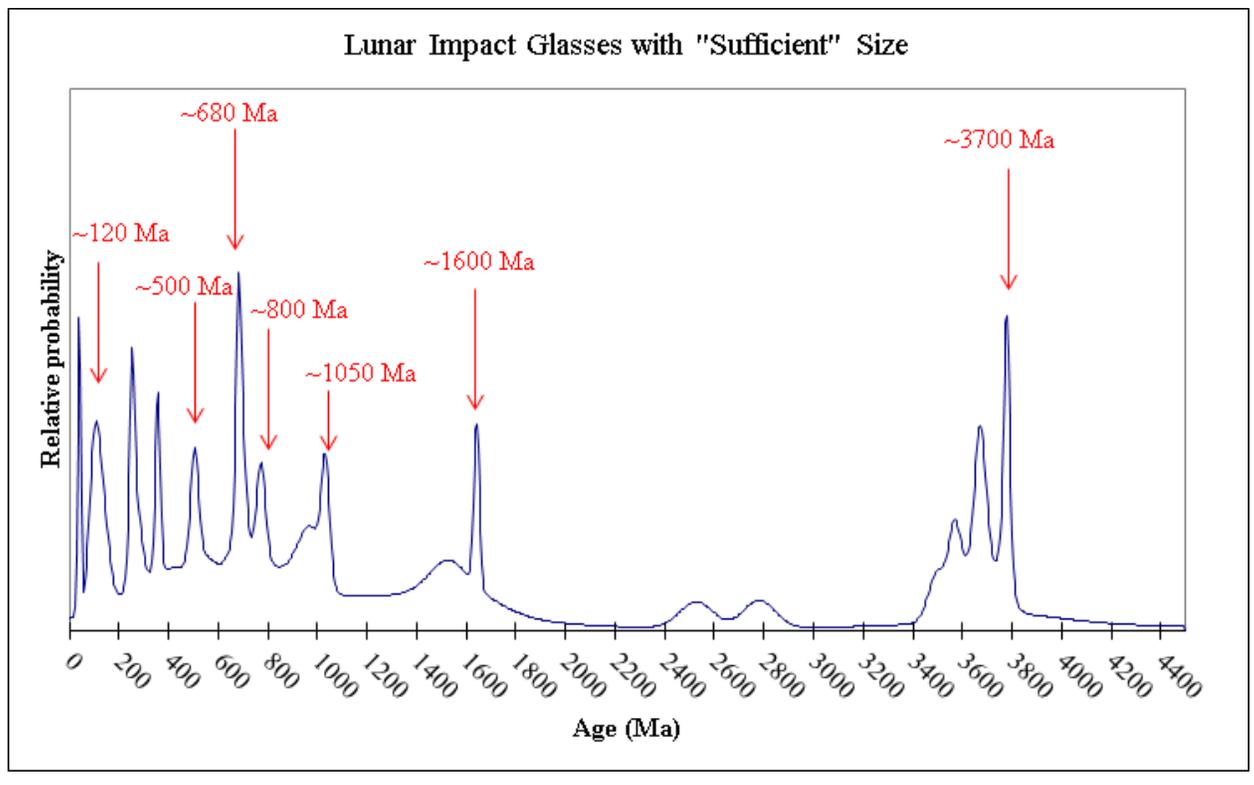

(b)

**Figure 8. (a)** Relative probability of impact ages occurring among ~100 impact glasses prior to application of the minimum required size criterion being applied. **(b)** Relative probability of impact ages occurring among the 48 impact glasses that satisfied the minimum required size criterion, and thus, with increased signal-to-noise ratio. Arrows indicate that at least three impact events were recorded in glass samples from more than two Apollo landing sites, implying at least a regional production and distribution of impact glasses with that age. The arrows identify impact episodes that have been documented elsewhere (e.g., ~500 Ma [Schmitz et al. 2001, 2003]; ~800 Ma [Zellner et al., 2009b; Swindle et al. 2009]; ~3700 Ma [Delano et al., 2007]), as well as others that may be statistically significant. Data in both figures are from this study (Table 1, Appendix B), Hui (2011), Zellner et al. (2009a,b), Delano et al. (2007), and Ryder et al. (1996). The scale on the y-axis is the same in both figures. Glasses that yielded no $^{40}$Ar/$^{39}$Ar ages ("ND"; Appendix B) are not included in either figure.

### 5.2.3 Diminished Preservation of Impact Glasses with Time

Following application of the minimum size criterion, Figure 7 displays a prominent decline in the frequency of all impact glasses with increasing age up to ~3500 Ma. This specific observation goes beyond the geologically short lifespans of impact glass spheres due to thermal strain, since impact glass shards show a decline in frequency with increasing age, too. A half-life of ~1000 Ma is indicated by that decline. While the trend in Figure 7 could be hypothesized as being caused by an increasing cratering rate during the last ~3500 Ma, we suspect that a more plausible cause of the trend is that lunar impact glasses gradually shatter into smaller pieces with time due to the thermal strain and impact-gardening of the lunar regolith.

### 5.3 Impact Glasses with $^{40}$Ar/$^{39}$Ar Ages >3500 Ma

Figure 7 shows 10 shards and spheres with ages of formation that are >3500 Ma forming a distinct age-frequency peak. These old impact glasses have been identified at the Apollo 14, 16, and 17 landing sites. The large compositional range (X(NBO) = 0.21-0.38) among these impact glasses (Figure 6) and the occurrence of three peaks in Figure 8b suggest that they are products of multiple impact melting events into compositionally diverse regions. While Culler et al. (2000) and Muller et al. (2001) also reported several peaks within that interval, it is well known from lunar sample analysis (Nyquist and Shih, 1992; Huneke, 1978; Papanastassiou et al., 1977; Turner, 1977) and photogeology (Hiesinger et al., 2000; Head 1976; Wilhelms and McCauley, 1971) that the Moon was undergoing extensive volcanism during that time in the form of crystalline mare basalts and picritic volcanic glasses. Therefore, in order to determine cratering rates, it is essential to distinguish between lunar volcanic glasses and lunar impact glasses, so that data from impact glasses only are plotted (as in Figures 6 and 7). Culler et al. (2000) and Muller et al. (2001), for example, did not chemically analyze their glasses prior to $^{40}$Ar/$^{39}$Ar dating, but rather assumed that volcanic glasses were not a significant component in their suite of Apollo 14 glasses. Delano (1988) observed that volcanic glasses were common (~50%) among the hundreds of glasses analyzed in Apollo 14 regolith breccias. Although a lower percentage of volcanic glasses was reported in Apollo 14 regoliths by the Apollo Soil Survey (1971) and Reid et al. (1973), the assumption by Culler et al. (2000) was flawed at some level. Consequently, the data reported by Culler et al. (2000) and Muller et al. (2001) are likely to be contaminated to some extent by ages of volcanic glasses, whereas the peaks in the current investigation within the age-interval 3500-3800 Ma (Figures 8a,b) are composed exclusively of ages from lunar impact glasses.

Among the >3500 Ma impact glasses in Figures 7 and 8b are the lmHKFM impact glasses (Delano et al., 2007), also known as 'basaltic andesitic' ('BA') glasses (Korotev et al., 2010; Zeigler et al., 2006). Those impact glasses, which are found most frequently at the Apollo 16 landing site, have a chemical composition that is exotic to the Apollo 16 site (Korotev et al., 2010; Delano et al., 2007; Zeigler et al., 2006) with X(NBO) = 0.21-0.24, and $^{40}Ar/^{39}Ar$ ages of 3730 ± 40 Ma (Delano et al., 2007). A potential source-crater of these lmHKFM glasses could be either Robertson (90 km diameter) or McLaughlin (80 km diameter), both of which occur in the Procellarum-KREEP terrain (as inferred by Korotev et al., 2010 and Zeigler et al., 2006) and have ages of 3700 ± 100 Ma (Kirchoff et al., 2013).

If, as previously discussed in Section 5.2.1, the gradual decline in the occurrence of lunar impact glasses with time is due largely to spontaneous shattering due to thermal strain, then the prominent occurrence of impact glasses with $^{40}Ar/^{39}Ar$ ages of 3500-3800 Ma (Figures 7,8b) requires an additional perspective. We suggest that those impact glasses could represent the lingering remnants of an initially large population of impact glasses generated during the tail end of the late heavy bombardment. The absence of lunar impact glasses with $^{40}Ar/^{39}Ar$ ages >3900 Ma could reflect (i) an increased rate of shattering of glasses during vigorous gardening of the regolith during the late heavy bombardment, as well as (ii) higher rates of diffusive Ar loss from impact glasses when the regolith had a steeper thermal gradient than the present one (Figure 2).

Since the lunar highlands surface has been dominated by feldspathic materials with X(NBO) ≤ 0.25 throughout most of the Moon's history, impact glasses derived from fusion of feldspathic highlands materials would have to be large (>1 cm; Figure 3) in order to preserve $^{40}Ar/^{39}Ar$ ages >3900 Ma (e.g., Imbrium impact event at 3934 ± 3 Ma; Merle et al., 2014). If, in addition, the lunar regolith was warmer at >3900 Ma (Nemchin et al., 2009), then the minimum required size of feldspathic impact glass with X(NBO) ≤ 0.25 would likely be >>1 cm (Figure 3) in order to yield reliable $^{40}Ar/^{39}Ar$ ages >3900 Ma. Since no such impact glasses have yet been identified in the current suite of lunar samples, lunar feldspathic impact glasses with $^{40}Ar/^{39}Ar$ ages >3900 Ma are likely to be exceptionally rare. Thus, $^{40}Ar/^{39}Ar$ dating of feldspathic lunar impact glasses is not likely to provide much information about very old episodes of lunar bombardment. Alternatively, if large impact basins, such as South Pole-Aitken, melted mafic lithologies (Hand, 2008; Hurwitz and Kring, 2014; Pieters et al., 2001, 2010) and produced glasses, then such impact glasses would have high values of X(NBO) and low Ar diffusivities compared to feldspathic glasses (Figure 3). Such as-yet-undiscovered impact glasses would have the potential of yielding reliable $^{40}Ar/^{39}Ar$ ages for impact events at >3900 Ma.

### 5.4 Lunar Impact Glasses and Biomolecular Clocks
With careful attention to chemical composition, size of sample, and exposure history, lunar impact glasses should be capable of providing important information about the bombardment history of the Earth-Moon system during at least the last ~3800 Ma. If, in addition to the Cretaceous/Tertiary mass extinction event (Alvarez et al., 1980), any other major biological events in Earth's biological history have been influenced by brief episodes of increased bombardment, then an important link might ultimately be found between the ages of lunar impact glasses and the timing of biological events inferred from biomolecular clocks (Knoll, 2014; Hedges and Kumar, 2009). With improved accuracy in the dating of lunar impact glasses and calibration of biomolecular clocks, the Moon may ultimately be recognized as a 'witness plate' for biologically important events (Delano et al., 2010).

## 5.5 Reporting Data
To allow the independent assessment of the quality of lunar impact glass data, future investigations should include morphological information (e.g., color, shape, size), geochemical composition (including analytical uncertainty in the measurements and X(NBO)), $^{40}Ar/^{39}Ar$ data (including 2σ-uncertainty in the ages), and an evaluation of whether or not the data set includes multiple glasses that may have formed in the same impact event (Figure 8b). In addition, when available, CRE ages, and inferred D(T,t) of the glass would be useful for application of the minimum size criterion for the measured exposure history; otherwise, an assumed exposure history, as described in the current study, would be required. Compositional data, including X(NBO) values, and ages for all of the glasses described herein are included in Table 1 and Appendices A, B and C.

# 6. Conclusions
We have analysed ~100 inclusion-free lunar impact glasses and provide geochemical and chronological data on 73 of them for the first time. Our findings are as follow: ***(i)*** Size, shape, chemical composition, and rates of diffusive loss of radiogenic $^{40}Ar$ are important for interpreting $^{40}Ar/^{39}Ar$ ages of lunar impact glasses. ***(ii)*** The age-distribution of lunar impact glass spherules (Figure 4) is dominated by ages <1000 Ma. In contrast to ancient lunar volcanic glasses that commonly occur as spherules, impact glass *spherules* may be prone to shattering into angular glass shards during impact gardening of the lunar regolith due to thermal stresses in those impact glasses acquired during quenching from hyperliquidus temperatures. If this inference is correct, $^{40}Ar/^{39}Ar$ age-distributions of lunar impact glass *spherules* would be intrinsically biased toward young ages and point misleadingly toward a recent increase in the impact flux. ***(iii)*** The accuracy of $^{40}Ar/^{39}Ar$ ages of lunar impact glasses is related to size and chemical composition. Based on the empirical results of this study and the experimental results of Gombosi et al. (2015), the retention of radiogenic $^{40}Ar$ in lunar impact glasses, and hence the reliability of $^{40}Ar/^{39}Ar$ ages, increases with physical size and increasing X(NBO) values of the glass sample. ***(iv)*** The age distribution of all impact glasses in Figure 7 and Figure 8b may reflect two distinct processes: diminished preservation of impact glasses with increasing age caused by shattering into smaller pieces during impact gardening of the regolith; and the preservation of a remnant population of impact glasses with ages >3500 Ma that survived from the tail end of the late heavy bombardment.

**Acknowledgments:** The authors thank Tim Swindle for his insight and guidance in interpreting $^{40}Ar/^{39}Ar$ ages and Clark Isachsen, Eric Olsen, and Fernando Barra for assistance with obtaining the argon data. Useful comments by Ryan Zeigler, Greg Herzog, Marc Norman, and an anonymous reviewer helped to improve the manuscript from an earlier version. Work by NEBZ was funded by the NASA LASER program grant #09-LASER09-0038 and by NSF Division of Astronomical Sciences grant #1008819. Work by JWD was supported by NASA Astrobiology grant #1079329-1-50310.

# References

Alvarez L. W., Alvarez W., Asaro F., and Michel, H. V. (1980) Extraterrestrial cause for the Cretaceous Tertiary Extincion. Science 208, 1095.

Apollo Soil Survey (1971) Apollo 14 - Nature and origin of rock types in soil from the Fra Mauro Formation. Earth Planet. Sci. Lett., 12, 49-54.

Arndt J., Englehardt W. V., Gonzalez-Cabeza I., and Meier B. (1984) Formation of Apollo 15 green glass beads. J. Geophys. Res. 89, C225-C232.

Bland P.A. (2005) The impact rate on Earth, Roy. Soc. Lon. Trans. A., 363, 2793-2810.

Culler T. S., Becker T. A., Muller R. A., and Renne P. R. (2000) Lunar impact history from 40Ar/39Ar dating of glass spherules. Science 287, 1785-1788.

Deino A. L. (2001) Users Manual for Mass Spec v. 5.02: Berkeley Geochronology Center Special Publication 1a, 119 p.

Delano J.W. (1975) The Apollo 16 mare component: Mare Nectaris. Proc. 6th Lunar Sci. Conf., 15-47.

Delano J.W. (1979) Apollo 15 green glass: Chemistry and possible origin. Geochim. et Cosmochim. Acta, Supplement 11, Proc. Lunar Planet. Sci. Conf. 10th, p. 275-300.

Delano J. W. (1986) Pristine lunar glasses: Criteria, data, and implications. Proc. 16th Lunar Planet. Sci. Conf., D201–D213.

Delano J. W. (1988) Apollo 14 regolith breccias: Different glass populations and their potential for charting space/time variations. Proc. 18th Lunar Planet. Sci. Conf., 59-65.

Delano J. W. (1991) Geochemical comparison of impact glasses from lunar meteorites ALHA81005 and MAC88105 and Apollo 16 regolith 64001. Geochim. Cosmochim. Acta 55, 3019-3029.

Delano J. W., Zellner N. E. B., and Swindle T. D. (2010) Lunar impact glasses and biomolecular clocks. Annual Meeting of the Lunar Exploration and Analysis Group, 3035.pdf. Lunar and Planetary Institute, Houston.

Delano J. W., Zellner N. E. B., Barra F., Olsen E., Swindle T. D., Tibbetts N. J., and Whittet D. C. B. (2007) An integrated approach to understanding Apollo 16 impact glasses: Chemistry, isotopes, and shape. Meteorit. Planet. Sci. 42, 6, 993-1004.

Eugster O. (2003) Cosmic-ray exposure ages of meteorites and lunar rocks and their significance. Chemie der Erde, 63, 3-30.

Eugster O., Grogler N., Eberhardt P. and Geiss J. (1979) Double drive tube 74001/2: History of the black and orange glass; Determination of a pre-exposure 3.7 AE ago by $^{136}$Xe/$^{235}$U dating. Proc. 10th Lunar Planet. Sci. Conf., 1351-1379

Farges F., Brown G. E., Jr., and Rehr J. J. (1996) Coordination chemistry of Ti(IV) in silicate glasses and melts: 1. XAFS study of titanium coordination in oxide model compounds. Geochim. Cosmochim. Acta 60, 3023-3038.

Fassett C. I. and Minton D. A. (2013) Impact bombardment of the terrestrial planets and the early history of the Solar System. Nature Geoscience 6 (doi: 10.1038/NGEO1841).

Gombosi D. J., Baldwin S. L., Watson E. B., Swindle T. D., Delano J. W., and Roberge W. G. (2015) Argon diffusion in Apollo 16 impact glass spherules: Implications for 40Ar/39Ar dating of lunar impact events. Geochim. Cosmochim. Acta, 148, 251-268.

Grier J. A. and McEwen A. S. (2001) The lunar record of impact cratering, in Accretion of Extraterrestrial Matter throughout Earth's History, (B. Peucker-Ehrenbrink and B. Schmitz, eds.), Chapter 22, 403-422. Kluwer Academic/Plenum Publishers, New York.

Grieve R. A. F. and Shoemaker E. M. (1994) The record of past impacts on Earth, in Hazards Due to Comets and Asteroids, ed. T. Gehrels, pp. 417-462, Univ. of Ariz. Press, Tucson.

Hand E. (2008) Planetary science: The hole at the bottom of the Moon. Nature, 453, 1160-1163.

Head J. W. III (1976) Lunar volcanism in space and time. Rev. Geophys., 14 (2), 265-300.

Hedges S. B. and Kumar S. (2009) The Timetree of Life. Oxford Univ. Press, New York. 551 pp.

Heiken G. H., McKay D. S., and Brown R. W. (1974) Lunar deposits of possible pyroclastic origin. Geochim. Cosmochim. Acta, 38, 1703-1718.

Hiesinger H., Jaumann R., Neukum G., and Head J. W. III (2000) Ages of mare basalts on the lunar nearside. Journ. Geophys. Res., 105 (E12), 29239-29275.

Hörz F. and Cintala M. (1997) Impact experiments related to the evolution of planetary regoliths. Meteorit. Planet. Sci. 32, 179-209.

Hudson G. B. (1981) Noble gas retention chronologies in the St. Severin meteorite. Ph.D. Dissertation, Washington University Department of Physics, St. Louis, MO.

Hui S. S. M. (2011) Microanalysis of impactors of the earth and impact products from the Moon: the origin and evolution of micrometeorites, and tracking bombardment history through chemical and radioisotopic memories of lunar impact spherules. Ph.D. Dissertation,

http://library.anu.edu.au/record=b2569918, The Australian National University, Canberra, Australia.

Hui S., Norman M. D., and Jourdan F. (2010) Tracking formation and transport of Apollo 16 lunar impact glasses through chemistry and dating. In Proc. 9th Australian Space Sci. Conf. Wayne Short and Iver Cairns, editors, p. 43-54 National Space Society of Australia Ltd., Sydney.

Huneke J. C. (1978) 40Ar-39Ar microanalysis of single 74220 glass balls and 72435 breccia clasts. Proc. Lunar Planet. Sci. Conf. 9th, 2345-2362.

Huneke J. C., Jesserberger E. K., Podosek F. A., and Wasserburg G. J. (1974) The age of metamorphisin of a highland breccia (65015) and a glimpse at the age of its protolith, in Lunar Science V, pp. 375-377, The Lunar Science Institute, Houston.

Hurwitz D. M. and Kring D. A. (2014) Differentiation of the South Pole-Aitken basin impact melt sheet: Implications for lunar exploration. Journ. Geophys. Res.: Planets, 119 (6), 1110-1133.

Husain L. and Schaeffer O. A. (1973) Lunar volcanism: Age of the glass in the Apollo 17 orange soil. Science, 180, 1358-1360.

Jourdan F. (2012) The 40Ar/39Ar dating technique applied to planetary science and terrestrial impacts. Aust. Journ. Earth Sci 59, 199-224.

Jourdan F. and Renne P. R. 2007. Age calibration of the Fish Canyon sanidine 40Ar/39Ar dating standard using primary K-Ar standards. Geochimica et Cosmochimica Acta 71:387–402.

Keller L. P. and McKay D. S. (1992) Micrometer-sized glass spheres in Apollo 16 soil 61181: Implications for impact volatilization and condensation. Proc. Lunar Planet. Sci. Conf. 22nd, 137-141.

Kirchoff M. R., Chapman C. R., Marchi S., Curtis K. M., Enke B., and Bottke W. F. (2013) Ages of large lunar impact craters and implications for bombardment during the Moon's middle age. Icarus 225, 325-341.

Knoll A. H. (2014) Paleobiological perspectives on early eukaryotic evolution, Cold Spring Harb. Perspect. Biol,, 6:a016121.

Korotev R. L. (1998) Concentrations of radioactive elements in lunar materials. J. Geophys. Res. 103, 1691–1701.

Korotev R. L. (2005) Lunar geochemistry as told by lunar meteorites. Chemie der Erde 65 (4), 297-346.

Korotev R. L., Zeigler R.A., and Floss C. (2010) On the origin of impact glass in the Apollo 16 regolith. Geochim. Cosmochim. Acta 74, 7362-7388.

Langseth M. G., Keihm S. J., and Peters K. (1976) Revised lunar heat-flow values. Proc. Lunar Sci. Conf. 7th, 3143-3171.

Lawson S. L. and Jakosky B. M. (1999) Brightness temperatures of the lunar surface: The Clementine long-wave infrared global data set. Lunar and Planetary Science Conf. 30th, 1892.pdf. Lunar and Planetary Institute, Houston, TX.

Lee S. K. (2011) Simplicity in melt densification in multicomponent magmatic reservoirs in Earth's interior revealed from multinuclear magnetic resonance. Proc. Natl. Acad. Sci. USA 108, 6847-6852.

LeFeuvre M. and Wieczorek M. A. (2011) Nonuniform cratering of the Moon and a revised crater chronology of the inner Solar System. Icarus 214, 1-20.

Levine J., Becker T. A., Muller R. A., and Renne P. R. (2005) 40Ar/39Ar dating of Apollo 12 impact spherules. Geophys. Res. Lett. 32, L15201, doi:10.1029/2005GL022874.

McDougall I. and Harrison T. M. (1999) Geochronology and Thermochronology by the $^{40}Ar/^{39}Ar$ Method. (2nd edition) Oxford University Press, New York. 269pp.

McEwen A. S., Moore J. M., and Shoemaker E. M. (1997) The Phanerozoic impact cratering rate: Evidence from the farside of the Moon. Journ. Geophys. Res., 102 (E4), 9231-9242.

Merle R. E., Nemchin A. A., Grange M. L., Whitehouse M. J., and Pidgeon R. T. (2014) High resolution U-Pb ages of Ca-phosphates in Apollo 14 breccias: Implications for the age of the Imbrium impact. Meteoritics & Planetary Science, 49, 2241-2251.

Mitchell J. K., Houston W. N., Scott R. F., Costes N. C., Carrier W. D. III, and Bromwell L. G. (1972) Mechanical properties of lunar soil: Density, porosity, cohesion, and angle of internal friction. Proc. Lunar Sci. Conf. 3rd 3, 3235-3253.

Morbidelli A., Marchi S., Bottke W. F., and Kring D. A. (2012) A sawtooth-like timeline for the first billion years of lunar bombardment. Earth Planet. Sci. Lett. 355-356, 144-151.

Muller R. A. (2002) Measurement of the lunar impact record for the past 3.5 b.y. and implications for the Nemesis theory, in Catastrophic Events and Mass Extinctions: Impacts and Beyond (C. Koeberl and K. G. MacLeod, eds.), Geol. Soc. Amer. Spec. Paper 356, 659-665.

Muller R. A., Becker T. A., Culler T. S., and Renne P. R. (2001) Solar system impact rates measured from lunar spherule ages, in Accretion of Extraterrestrial Matter throughout Earth's History (B. Peucker-Ehrenbrink and B. Schmitz, eds.), Chapter 22, 447-462. Kluwer Academic/Plenum Publishers, New York.

Mysen B. O. and Richet P. (2005) Silicate Glasses and Melts: Properties and Structure. Elsevier, Amsterdam. 560pp.

NRC, The National Research Council (2007) The Scientific Context for Exploration of the Moon, The National Academies Press.

Nemchin A. A., Pidgeon R. T., Healy D., Grange M. L., Whitehouse M. J., and Vaughn J. (2009) The comparative behavior of apatite-zircon U-Pb systems in Apollo 14 breccias: Implications for the thermal history of the Fra Mauro Formation. Meteoritics & Planetary Science, 44, (11), 1717-1734.

Neukum G., Ivanov B. A., and Hartmann W. K. (2001) Cratering records in the inner Solar System in relation to the lunar reference system. Space Sci. Rev. 96, 55-86.

Norman M. D., Adena K. J. D., and Christy A. G. (2012) Provenance and Pb isotopic ages of volcanic and impact glasses from the Apollo 17 landing site on the Moon. Aust. Journ. Earth Sci. 59, 291-306.

Norman M.D., Hui S., and Adena K. (2011) The lunar impact record: Greatest hits and one hit wonders. In Proc. 10th Australian Space Sci. Conf., Canberra. 26 - 29 September, Published by the National Space Society of Australia Ltd, eds. W. Short and I. Cairns, ISBN 13: 978-0-9775740-5-6.

Nyquist L. E. and Shih C.-Y. (1992) The isotopic record of lunar volcanism. Geochim. Cosmochim. Acta, 56 (6), 2213-2234.

Papanastassiou D. A., DePaolo D. J., and Wasserburg G. J. (1977) Rb-Sr and Sm-Nd chronology and genealogy of mare basalts from the Sea of Tranquillity. Proc. Lunar Planet. Sci. Conf. 8th, 1639-1672. Pergamon Press, New York.

Pieters C., Head J. W. III, Gaddis L., Jolliff B., and Duke D. (2001) Rock types of South Pole-Aitken basin and extent of basaltic volcanism. Journ. Geophys. Res., 106 (E11), 28,001-28,022.

Pieters C. M., Ohtake M., Haruyama J., Jolliff B. L., Gaddis L. R., Petro N. E., Klima R. L., and Head J. W. III (2010) Implications of the distinctive Mafic Mound in central SPA. Amer. Geophys. Union, Annual Fall Meeting, abstract #P43A-04.

Podosek F. A. and Huneke J. C. (1973) Argon in Apollo 15 green glass spherules (15426): 40Ar-39Ar age and trapped argon. Earth Planet. Sci. Lett., 19 (4), 413-421.

Prettyman T. H., Hagerty J. J., Elphic R. C., Feldman W. C., Lawrence D. J., McKinney G. W., and Vaniman D. T. (2006) Elemental composition of the lunar surface: Analysis of gamma ray spectroscopy data from Lunar Prospector. Journ. Geophys. Res., 111 (E12). doi: 10.1029/2005JE002656

Reid A. M., Warner J., Ridley W. I., Johnston D. A., Harmon R. S., Jakes P., and Brown R. W. (1972a) The major element compositions of lunar rocks as inferred from glass compositions in the lunar soils. Proc. 3rd Lun. Sci. Conf., 363-378.

Reid A. M., Ridley W. I., Harmon R. S., Warner J., Brett R., Jakes P., and Brown R. W. (1972b) Highly aluminous glasses in lunar soils and the nature of the lunar highlands. Geochim. Cosmochim. Acta 36, 903-912.

Reid A. M., Ridley W. I., Harmon R. S., and Jakes P. (1973) Major element chemistry of glasses in Apollo 14 soil 14156. Geochim. Cosmochim. Acta, 37 (3), 695-699.

Renne P. R., Mundil R., Balco G., Min K., and Ludwig K. (2010) Joint determination of 40K decay constants and 40Ar*/40K for the Fish Canyon sanidine standard, and improved accuracy for 40ar/49Ar geochronology, Geochim. Cosmochim. Acta, 75, 5097-5100.

Rhodes J. M., Rodgers K. V., Shih C., Bansal B. M., Nyquist L. E., Wiesmann H., and Hubbard N. J. (1974) The relationships between geology and soil chemistry at the Apollo 17 landing site. Proc. 5th Lunar Sci. Conf., 1097-1117.

Ryder G., Delano J. W., Warren P. H., Kallemeyn G. W., and Dalrymple G. B. (1996) A glass spherule of questionable impact origin from the Apollo 15 landing site: Unique target mare basalt. Geochim. Cosmochim. Acta 60, 693-710.

Ryder G., Koeberl C., and Mojzsis S. (2000) Heavy bombardment of the Earth at ~3.85 Ga: The search for petrographic and geochemical evidence. In Origin of the Earth and Moon (R. Canup and K. Righter, eds.), 475-492, University of Arizona Press, Tucson, 555 pp.

Schmitz B., Häggström T., and Tassinari M. (2003) Sediment-Dispersed Extraterrestrial Chromite Traces a Major Asteroid Disruption Event, Science, 300, 961-964.

Schmitz B., Tassinari M., and Peucker-Ehrenbrink B. (2001) A rain of ordinary chondritic meteorites in the early Ordovician, Earth Plan. Sci. Lett., 194, 1-15.

Shoemaker E. M. (1983) Asteroid and comet bombardment of the Earth. Ann. Rev. Earth Planet. Sci. 11, 461-494.

Smith J. V. (1974) Lunar Mineralogy: A heavenly detective story Presidential Address. Amer. Mineral. 59, 231-243.

Spangler R. R., Warasila R., and Delano J. W. (1984) $^{39}$Ar-$^{40}$Ar ages for the Apollo 15 green and yellow volcanic glasses. Proc. 14$^{th}$ Lunar Planet. Sci. Conf., J. Geophys. Res., 89 (Supplement), B487-B-497.

Steele A. M., Colson R. O., Korotev R. L., and Haskin L. A. (1992) Apollo 15 green glass – Compositional distribution and petrogenesis. Geochim. Cosmochim. Acta 56 (11), 4075-4090.

Steiger R. H. and Jäger E. (1977) Subcommission on geochronology: Convention on the use of decay constants in geo- and cosmochronology, Earth Planet. Sci. Lett. 36, 359-362.

Stöffler D. and Ryder G. (2001) Stratigraphy and isotope ages of lunar geologic units: Chronological standard for the inner Solar System. Space Sci. Rev. 96, 9-54.

Stöffler D., Ryder G., Ivanov B. A., Artemieva N. A., Cintala M. J., and Grieve R. A. F. (2006) Cratering history and lunar chronology. Rev. Mineral. Geochem. 60, 519-596.

Swindle T. D., Isachsen C. E., Weirich J. R., and Kring D. A. (2009) 40Ar-39Ar ages of H-chondrite impact melt breccias. Met. Planet. Sci. 44, 747–762.

Sutton S. R., Jones K. W., Gordon B., Rivers M. L., Bajt S., and Smith J. V. (1993) Reduced chromium in olivine grains from lunar basalt 15555 – X-ray Absorption Near Edge Structure (XANES). Geochim. Cosmochim. Acta 57, 641-648.

Symes S. J. K., Sears D. W. G., Akridge D. G., Huang S., and Benoit P. H. (1998) The crystalline lunar spherules: Their formation and implications for the origin of meteoritic chondrules. Meteoritics and Planetary Science 33, 13-29.

Taylor G. J. (2009) Ancient lunar crust: Origin, composition, and implications. Elements, 5 (1), 17-22.

Tera F., Papanastassious D. A., and Wasserburg G. J. (1974) Isotopic evidence for a terminal lunar cataclysm. Earth Planet. Sci. Lett. 22, 1-21.

Turner G. (1977) Potassium-argon chronology of the Moon. Phys. Chem. Earth, 10 (3), 145-195.

Ulrich D. R. (1974) Study of recrystallization and devitrification of lunar glass. NASA-CR-134307, 58 pp.

Vasavada A. R., Bandfield J. L., Greenhagen B. T., Hayne P. O., Siegler M. A., Williams J.-P., and Paige D. A. (2012) Lunar equatorial surface temperatures and regolith properties from the Diviner Lunar Radiometer Experiment. J. Geophys. Res. 117, E00H81, doi:10.1029/2011JE003987.

Weirich J. R. (2011) Improvements to Argon-Argon Dating of Extraterrestrial Materials. Ph.D. Dissertation, The University of Arizona, Tucson, Arizona.

Wilhelms D. E. and McCauley J. F. (1971) Geologic map of the near side of the Moon. I-703. U.S. Geol. Survey, Washington, D.C.

Wu Y., Xue B., Zhao B., Lucey P., Chen J., Xu X., Li C., and Ouyang Z. (2012) Global estimates of lunar iron and titanium contents from the Chang'E-1 IIM data. Journ. Geophys. Res., 117 (E2). doi: 10.1029/2011JE003879

Wünnemann K., Collins G. S., and Osinski G. R. (2008) Numerical modelling of impact melt production in porous rocks. Earth Planet. Sci. Lett. 269, 530-539.

30Zeigler R. A., Haskin L. A., Korotev R. L., Jolliff B. L., and Gillis J. J. (2003) The Apollo 16 mare component: Petrography, geochemistry, and provenance. Lunar Planet. Sci. Conf.-XXXIV, 1454.pdf. Lunar and Planetary Institute, Houston.

Zeigler R.A., Korotev R. L., Jolliff B. L., Haskin L. A., and Floss C. (2006) The geochemistry and provenance of Apollo 16 mafic glasses. Geochim. Cosmochim. Acta 70, 6050-6067.

Zellner N. E. B., Spudis P. D., Delano J. W. and Whittet D. C. B. (2002) Impact glasses from the Apollo 14 landing site and implications for regional geology, J. Geophys. Res. 107 (E11), 5102, doi:10.1029/2001JE001800.

Zellner N. E. B, Delano J. W., Swindle T. D., Barra F., Olsen E., and Whittet D. C. B. (2009a) Apollo 17 regolith, 71501,262: A record of impact events and mare volcanism in lunar glasses. Meteorit. Planet. Sci. 44(6), 839-852.

Zellner N. E. B, Delano J. W., Swindle T. D., Barra F., Olsen E., and Whittet D. C. B. (2009b) Evidence from 40Ar/39Ar ages of lunar impact glasses for an increase in the impact rate ~800 Ma Ago. Geochim. Cosmochim. Acta 73, 4590-4597, 10.1016/j.gca.2009.04.027.